\documentclass[prl,twocolumn]{revtex4}
\usepackage{graphicx}
\usepackage{epstopdf}
\usepackage{amsmath,amssymb}
\usepackage{yfonts,mathrsfs,accents}

\newcommand\Lie{\mathscr{L}}

\newcommand\grad\nabla

\newcommand\dual{\mathord{*}}
\newcommand\ed{\mathrm{d}}

\newcommand\KL{\Delta_K}
\newcommand\oKL{\ring\Delta_K}
\newcommand\sKLs{\text{\setbox0\hbox{$\ring\Delta$}%
	\hbox to \wd0{\footnotesize\hss$\Delta$}\hskip-\wd0\box0}\vphantom{\Delta}_K}

\makeatletter

\def\nml#1{{\hbox{$\left#1\vbox to5.5\p@{}\right.\n@space$}}}

\newcommand\abs{\defaultoptions{\@dirac}{[\nml][||]}}
\newcommand\bra{\defaultoptions{\@dirac}{[\nml][<|]}}
\newcommand\ket{\defaultoptions{\@dirac}{[\nml][|>]}}
\newcommand\norm{\defaultoptions{\@dirac}{[\nml][\|\|]}}
\long\def\@dirac[#1][#2#3]#4{\mathopen#1#2 #4 \mathclose#1#3}

\newcommand\braket{\defaultoptions{\@braket}{[\nml][<>][|]}}
\long\def\@braket[#1][#2#3][#4]#5#6{\@dirac[#1][#2#3]{#5\mathopen#1#4#6}}

\newcommand\expect{\defaultoptions{\@expect}{[\nil][\nml][<>][||]}}
\long\def\@expect[#1][#2][#3#4][#5#6]#7{%
	\ifx#1\nil  \@dirac[#2][#3#4]{#7}  \else  \matel[#2][#3#4][#5#6]{#1}{#7}{#1}  \fi}

\newcommand\comm{\defaultoptions{\@iprod}{[\nml][\lbrack\rbrack]}}
\newcommand\iprod{\defaultoptions{\@iprod}{[\nml][<>]}}
\long\def\@iprod[#1][#2#3]#4#5{\@dirac[#1][#2#3]{#4,#5}}

\newcommand\matel{\defaultoptions{\@matel}{[\nml][\langle\rangle][||]}}
\long\def\@matel[#1][#2#3][#4#5]#6#7#8{\@dirac[#1][#2#3]{#6\@dirac[#1][#4#5]{#7}#8}}

\newcommand\defaultoptions[2]{\@defaultoptions{#1}{}{#2}}
\def\@defaultoptions#1#2#3{\@ifnextchar[{\@@defaultoptions{#1}{#2}#3\@nil}{#1#2#3}}
\def\@@defaultoptions#1#2[#3]#4\@nil[#5]{\@defaultoptions{#1}{#2[#5]}{#4}}

\newcommand\stsig[1]{(\@stsig#1\relax\@nil)}
\def\@stsig#1#2#3\@nil{\mathord{#1}  \ifx#2\relax\else  \@stsig#2#3\@nil  \fi}

\makeatother

\begin{document}

\title{Approximate Killing Fields as an Eigenvalue Problem}
\author{Christopher Beetle}
\affiliation{Department of Physics, Florida Atlantic University, Boca Raton, Florida, 33431}
\date{July 13, 2007}

\begin{abstract}
Approximate Killing vector fields are expected to help define physically meaningful spins for non-symmetric black holes in general relativity.  However, it is not obvious how such fields should be defined geometrically.  This paper relates a definition suggested recently by Cook and Whiting to an older proposal by Matzner, which seems to have been overlooked in the recent literature.  It also describes how to calculate approximate Killing fields based on these proposals using an efficient scheme that could be of immediate practical use in numerical relativity.
\end{abstract}

\maketitle

Spacetime symmetries are essential for defining physically important conserved quantities such as energy and angular momentum in general relativity.  For example, when a vacuum spacetime admits a rotational symmetry generated by a Killing vector field $\varphi^a$, then a Komar-type integral \cite{K:CCL, W:GR} over an arbitrary 2-sphere in spacetime can reproduce the physically well-defined angular momentum measured at infinity.  When spacetime is axi-symmetric, but not vacuum, the difference between these integrals for a pair of different 2-spheres is precisely the ordinary angular momentum computed from the stress-energy of matter in the intervening space.  If spacetime is \textit{not} axi-symmetric, however, such formulae become rather ambiguous.  They depend not only on the 2-sphere $S$ over which one integrates, but also on the vector field $\varphi^a$ used to define the integrand.

These difficulties can be partially avoided under physically favorable conditions.  For instance, angular momentum is well-defined at infinity in asymptotically flat spacetimes (see \cite{S:QLR} for a recent review), or on an appropriate isolated horizon \cite{ABL:IHJ, AK:IDH} modeling an isolated, non-dynamical black hole in a spacetime that may describe interesting dynamics in other regions.  Essentially, these techniques identify preferred 2-spheres $S$ (infinity, horizon, etc.) over which to integrate, and thereby reduce the ambiguity in defining the angular momentum.  The resulting quasi-local formulae have the general Brown--York \cite{BY:QLE} form 
\begin{equation}\label{BYJ}
	J[\varphi] := \frac{1}{8\pi G}\, \oint_S\, \varphi^a\, K_{ab}\, \ed S^b, 
\end{equation}
where $S$ is a (perhaps preferred) 2-sphere, $K_{ab}$ is the extrinsic curvature of a spatial slice $\Sigma$ containing it, $\varphi^a$ is a vector field on $S$, and $\ed S^b$ is the area element of $S$ within $\Sigma$.  The basic problem remains: the vector field $\varphi^a$ is arbitrary unless $S$ has an \textit{intrinsic} symmetry that can be used to select it.  (Now, at least, that symmetry need not extend into the bulk of spacetime.)

The horizons of black holes resulting from numerical simulations of astrophysical processes generally have no symmetry of any kind and therefore, seemingly, no preferred vector field $\varphi^a$.  The problem is not that such surfaces have \textit{no} reasonable definition for the angular momentum, but rather that they have \textit{infinitely many}.  There is one for every vector field $\varphi^a$ tangent to the horizon.  What is needed is a technique to pick a preferred vector field, and the obvious thing to do is to seek a $\varphi^a$ that, in some sense, is as close as possible to a Killing field, even if none is present.  This leads intuitively to the idea of an \textit{approximate} Killing field.

Motivated by the general issues discussed above, several groups have recently proposed elegant definitions of approximate Killing fields on 2-spheres.  These include schemes based on Killing transport \cite{DKSS:KT}, conformal Killing vectors \cite{CCGP:CKV}, and most recently a minimization scheme by Cook and Whiting \cite{CW:S2}.  This paper revives an older approach \cite{M:ASS} due to Matzner based on solving an eigenvalue problem.  It also suggests a novel adaptation of Matzner's approach to the specific problem of computing a preferred angular momentum for black holes in numerical relativity, and elucidates the intimate relationship between this scheme and that of Cook and Whiting.


Let us begin with a brief review of Matzner's definition \cite{M:ASS} of an approximate Killing field on a compact manifold $M$ of dimension $n$ equipped with a Riemannian metric $g_{ab}$.  A continuous symmetry of $g_{ab}$ is generated by a Killing vector field $\xi^a$ satisfying 
\begin{equation}\label{Keq}
	\Lie_\xi\, g_{ab} = 2\, \grad_{(a}\, \xi_{b)} = 0 
\end{equation}
throughout $M$, where $\Lie_\xi$ denotes the Lie derivative along $\xi^a$ and $\grad_a$ is the unique torsion-free connection on $M$ determined by $g_{ab}$.  Taking a divergence, we see that any geometry with a continuous symmetry admits at least one non-trivial solution to the equation 
\begin{equation}\label{Kop}
	-2\, \grad_b\, \grad^{(b}\, \xi^{a)} = 0.
\end{equation}
In principle, even when the geometry is \textit{not} symmetric, we are still free to seek solutions for this equation.  But generically we will not find any.

Consider the eigenvalue problem 
\begin{equation}\label{Kevp}
	\KL\, \xi^a := -2\, \grad_b\, \grad^{(b}\, \xi^{a)} = \kappa\xi^a
\end{equation}
on a generic geometry.  The operator $\Delta_K$ appearing here arises naturally in the transverse decomposition of symmetric tensor fields on Riemannian manifolds \cite{Y:DST}.  A related operator, denoted $\Delta_L$, arises in the same way from the \textit{conformal} Killing equation, and plays a similar role in the transverse-\textit{traceless} decomposition of such fields.  Its application to the initial-data problem in general relativity is very well known indeed \cite{YO:IVP1}.

Eq.~(\ref{Kevp}) of course admits solutions $\xi^a$ only for a certain spectrum of eigenvalues $\kappa$, and zero may or may not be among these.  Matzner \cite{M:ASS} establishes the following four results: the spectrum of eigenvalues $\kappa$ of Eq.~(\ref{Kevp}) on a compact manifold is (a) discrete, (b) non-negative, (c) corresponds to a complete set of \textit{real} vector eigenfields $\xi^a$, and (d) contains $\kappa = 0$ if and only if the corresponding eigenfield $\xi^a$ is a genuine Killing field.  That is, the zero eigenspace of Eq.~(\ref{Kevp}) is precisely the finite-dimensional vector space of Killing \textit{fields} of the metric $g_{ab}$ on $M$.  Therefore, Eq.~(\ref{Kop}) admits no solution if the metric $g_{ab}$ on $M$ has no continuous symmetries, as claimed above.  However, we assert that \textit{the best approximation to a Killing field on a manifold with no actual symmetry is the unique vector eigenfield $\xi^a$ of Eq.~\ref{Kevp} with the minimum  eigenvalue $\kappa > 0$}.  This is Matzner's definition of an approximate Killing field, and it has several desirable features.  It exists generically, reduces to the correct answer when symmetries do exist, and, like a true Killing field on a symmetric manifold, is naturally defined only up to an overall (i.e., constant over $M$) scaling.

Like any eigenvalue problem, Eq.~(\ref{Kevp}) admits a variational formulation.  Recall the natural $L^2$ inner product 
\begin{equation}\label{ipdef}
	\iprod{\zeta}{\xi} := \oint_M\, \overline{\zeta_a}\, \xi^a\, \epsilon
\end{equation}
on the space of (complex) vector fields over $M$.  Here, $\epsilon$ denotes the canonical $n$-form volume element induced on $M$ by the metric $g_{ab}$.  We minimize the quadratic form 
\begin{equation}\label{Kkap}
\begin{aligned}
	Q_K[\xi; \kappa) :={}& \tfrac{1}{2}\, \iprod{\xi}{\KL\, \xi} - \tfrac{1}{2}\kappa\, \bigl( \iprod{\xi}{\xi} - 1 \bigr), 
\end{aligned}
\end{equation}
where $\kappa$ is constant over $M$ and here plays the role of a Lagrange multiplier.  Minimizing this functional produces the Euler--Lagrange equations 
\begin{equation}\label{KEL}
	\KL\, \xi^a = \kappa\xi^a \quad\mbox{and}\quad \iprod{\xi}{\xi} = 1, 
\end{equation}
the solutions of which are clearly the vector eigenfields of Eq.~(\ref{Kevp}), normalized to unity in Hilbert space.

Many variational problems are solved by initially solving the first, differential equation in Eq.~(\ref{KEL}) for $\xi^a$ as a function of an arbitrary Lagrange multiplier $\kappa$, and then using that result in the second, algebraic equation to impose the constraint and determine $\kappa$.  This does not happen for Eq.~(\ref{KEL}) because the differential equation is linear in $\xi^a$, and therefore leaves the overall scaling of $\xi^a$ undetermined.  The second equation serves only to fix that scaling, and cannot also determine $\kappa$.  The Lagrange multiplier therefore must be fixed when we solve the \textit{first} equation; for general $\kappa$, no solution exists.  This is hardly surprising since of course only true eigenvalues $\kappa$ allow us to solve Eq.~(\ref{Kevp}) for $\xi^a$.  However, it does make an approach to Matzner's eigenvalue problem via a variational principle like Eq.~(\ref{Kkap}) rather complicated.  There is no algebraic equation to determine the Lagrange multiplier.  Indeed, $\kappa$ is determined in this problem precisely by the condition that it be an eigenvalue of $\KL$, and there is no algebraic equation giving these.  Minimizing $Q_K[\xi; \kappa)$ in Eq.~(\ref{Kkap}) by solving the associated Euler--Lagrange equations is neither easier nor harder than solving the eigenvalue problem in Eq.~(\ref{Kevp}).

Cook and Whiting's recent definition \cite{CW:S2} of an approximate Killing field uses a variational principle based on a quadratic form closely related to that of Eq.~(\ref{Kkap}).  However, it differs in a two important details.  First, it focuses on the case where $M \simeq S$ is topologically a 2-sphere, and restricts $\xi^a$ to be area-preserving: 
\begin{equation}\label{arpre}
	\Lie_\xi\, \epsilon_{ab} = (\grad_c\, \xi^c)\, \epsilon_{ab} = 0.
\end{equation}
The motivation for this restriction arises from the technical details of an eventual application to calculating the angular momentum of a non-symmetric black hole \cite{ABL:IHJ}.  Second, it is based on a non-standard inner product 
\begin{equation}\label{ipR}
	\iprod{\zeta}{\xi}_R := \iprod{\zeta}{R\, \xi} = \oint_S\, \overline{\zeta_a}\, \xi^a\, R\, \epsilon.
\end{equation}
These choices change the form, but not the basic content, of the resulting equations.  They still describe a sort of self-adjoint eigenvalue problem.

To restrict to area-preserving vector fields, it is easiest simply to recall that any divergenceless vector field $\ring\xi^a$ on a 2-sphere topology is described by a unique scalar potential $\Theta$ such that 
\begin{equation}\label{xipot}
	\ring\xi^a = (\dual\ed\Theta)^a := -\epsilon^{ab}\, \grad_b\, \Theta \quad\mbox{and}\quad \oint_S\, \Theta\, \epsilon = 0.
\end{equation}
Now consider the restricted eigenvalue problem
\begin{equation}\label{rKevp}
	\oKL\, \ring\xi^a := \bigl( \ring P\, \KL \ring P \bigr)\, \ring\xi^a = \ring\kappa\, \ring\xi^a, 
\end{equation}
where $\ring P$ denotes the orthogonal projection onto the subspace of area-preserving vector fields within the Hilbert space of Eq.~(\ref{ipdef}).  A given $\ring\xi^a = (\dual\ed\Theta)^a$ solves this equation if and only if, for all other $\ring\zeta^a = (\dual\ed\Phi)^a$, we have 
\begin{equation}\label{rKips}
	\iprod{\dual\ed\Phi}{\KL\, \dual\ed\Theta} = \ring\kappa\, \iprod{\dual\ed\Phi}{\dual\ed\Theta}.
\end{equation}
Integrating both sides by parts, and using positivity of the standard $L^2$ inner product of scalar functions over $S$, we find that Eq.~(\ref{rKevp}) is equivalent to 
\begin{equation}\label{Ldef}
	\sKLs\, \Theta := \Delta\!^2\, \Theta + \grad^a\, (R\, \grad_a\, \Theta) = \ring\kappa\, \Delta\, \Theta, 
\end{equation}
where $\Delta := -\grad^a\, \grad_a$ denotes the standard scalar Laplacian.  We have shown that the scalar functions $\Theta$ solving Eq.~(\ref{Ldef}) generate, via Eq.~(\ref{xipot}), solutions $\ring\xi^a$ of the restricted eigenvalue problem of Eq.~(\ref{rKevp}).  Because the projection $\ring P$ does not typically commute with $\KL$, these $\ring\xi^a$ do not generally solve Eq.~(\ref{Kevp}), and the restricted eigenvalues $\ring\kappa$ are generally distinct from the eigenvalues $\kappa$ in the full Hilbert space.  In fact, we should generally expect that $\ring\kappa_{\mathrm{min}} > \kappa_{\mathrm{min}}$.  However, the area-preserving vector eigenfield corresponding to this minimum restricted eigenvalue can also be considered a best approximation to a Killing field, albeit within a restricted class.

To recover the Cook--Whiting approximate Killing field, we must repeat the previous calculation in the inner product of Eq.~(\ref{ipR}).  The operator $\KL$ is then no longer self-adjoint, but $R^{-1}$ times $\KL$ is.  Accordingly, we seek vector fields $\ring\xi_R^a = (\dual\ed\Theta)^a$ satisfying 
\begin{equation}\label{rKipR}
	\iprod{\dual\ed\Phi}{R^{-1}\, \KL\, \dual\ed\Theta}_R = \ring\kappa_R\, \iprod{\dual\ed\Phi}{\dual\ed\Theta}_R
\end{equation}
for all $\ring\zeta^a = (\dual\ed\Phi)^a$.  Integrating by parts, and once again invoking positivity of the \textit{standard} inner product of scalar functions, we find that Eq.~(\ref{rKipR}) is equivalent to 
\begin{equation}\label{CWevp}
	\sKLs\, \Theta = -\ring\kappa_R\, \grad^a\, (R\, \grad_a\, \Theta).
\end{equation}
Although the notation here differs slightly, this is precisely the Euler--Lagrange equation that Cook and Whiting find \cite{CW:S2} by minimizing a quadratic form similar to Eq.~(\ref{Kkap}).  Once again, the solutions $(\ring\xi^a_R, \ring\kappa_R)$ of this eigenvalue problem generally differ from the solutions $(\xi^a, \kappa)$ of Eq.~(\ref{Kevp}) and from the solutions $(\ring\xi^a, \ring\kappa)$ of Eq.~(\ref{rKevp}).

Let us now make two technical comments.  First, any constant function $\Theta = c$ will give zero on both sides of Eqs.~(\ref{rKevp}) and (\ref{CWevp}) for all values of $\ring\kappa$ or $\ring\kappa_R$, respectively.  These are spurious solutions.  They arise only because we have used potentials to describe the subspace of area-preserving vector fields.  These solutions are ruled out by the the second condition in Eq.~(\ref{xipot}), which makes the correspondence between $\ring\xi^a$ and $\Theta$ an isomorphism.

Second, the Cook--Whiting inner product in Eq~(\ref{ipR}) looks a little odd, but it is not immediately clear whether there is anything technically wrong with it.  There certainly can be problems.  Recall that the scalar curvature in two dimensions varies as 
\begin{equation}\label{Rdot}
	\delta\, {}^2\! R = \grad^b\, \grad_{[a}\, \delta g_{b]}{}^a - \tfrac{1}{2}\, {}^2\!R\, \delta g_a{}^a
\end{equation}
under a perturbation $\delta g_{ab}$ of the metric.  If this perturbation varies sufficiently rapidly over $S$, then the first term here can easily dominate the second, as well as the unperturbed, background value ${}^2\!R$.  The result is that a generic spherical geometry, even if perturbatively close to a round sphere in the sense that $\delta g_{ab}$ has small amplitude, can have regions of negative scalar curvature.  (This is intuitively obvious if we imagine ``pinching'' the surface of a round sphere to create a small, saddle-shaped region of negative curvature.)  On such geometries, the ``inner product'' of Eq.~(\ref{ipR}) is not positive-definite, and does not define a Hilbert space.  However, in the space of all spherical geometries, there should be some finite region of ``sufficiently smooth'' perturbations of the round sphere for which the total scalar curvature remains everywhere positive.  In this region, there is no obvious problem with the Cook--Whiting scheme, but nothing particular to recommend it either.  The question could presumably be settled \cite{num} by comparing qualitative features of the approximate Killing fields computed from Eqs.~(\ref{rKevp}) and (\ref{CWevp}).

Matzner's eigenvalue definition of an approximate Killing field is unambiguous, universally applicable, and reproduces the usual Killing fields on a symmetric manifold.  But it is not necessarily efficient in practice.  Indeed, it would prohibitively expensive to solve any of the eigenvalue problems in Eqs.~(\ref{Kevp}), (\ref{rKevp}) or (\ref{CWevp}) on the apparent horizon at every moment of time of a black hole in a numerical simulation.  Fortunately, however, there is a simple approximation to speed the process up on a generic geometry.  This approximation is based on the Rayleigh--Ritz method \cite{MW:MMP}, and works so long as we only want to find the \textit{lowest} eigenvalue and the corresponding vector eigenfield.

Consider the Rayleigh--Ritz functional 
\begin{equation}\label{Khat}
	F[\xi] := \frac{\iprod{\xi}{\KL\, \xi}}{\iprod{\xi}{\xi}} 
		= \frac{\displaystyle 2\, \oint_M\, \grad_{(a}\, \xi_{b)} \cdot \grad^{(a}\, \xi^{b)} \cdot \epsilon}
			{\displaystyle \oint_M\, \xi_a\, \xi^a\, \epsilon}
\end{equation}
on the full Hilbert space of Eq.~(\ref{ipdef}), with the zero vector removed.  The local extrema of Eq.~(\ref{Khat}) occur when $\xi^a$ is a vector eigenfield of $\KL$, and the value of $F[\xi]$ at each such extremum is the corresponding eigenvalue $\kappa$.  Note that the numerator here, which arises via integration by parts of the second-order operator $\KL$ in Eq.~(\ref{Kevp}), is precisely one half the square integral of $\Lie_\xi\, g_{ab}$ from Eq.~(\ref{Keq}).  Thus, among all vector fields with fixed $L^2$-norm on $S$, diffeomorphisms along the approximate Killing field modify the metric least in a quantifiable, $L^2$ sense.

It is still not practicable to find the genuine absolute minimum of Eq.~(\ref{Khat}) on the computer, which of course would yield Matzner's approximate Killing field.  But one can \textit{approximate} that minimum by minimizing $F[\xi]$ within an appropriate space of trial vector fields.  This idea is familiar from elementary quantum mechanics, where just such a variational principle is routinely used to approximate the ground-state wave-function of a complicated system.  Unless the subspace of trial fields one chooses happens to be orthogonal, or nearly so, to the true minimum $\xi^a_{\mathrm{true}}$ of $F[\xi]$ in all of Hilbert space, the dominant component of the minimizing trial field $\xi^a_{\mathrm{trial}}$ should lie along $\xi^a_{\mathrm{true}}$ in Hilbert space.  Most randomly-chosen trial spaces will not be orthogonal to $\xi^a_{\mathrm{true}}$.  This idea allows us to approximate Matzner's approximate Killing field.

There is a natural candidate for the trial space of vector fields in which to minimize Eq.~(\ref{Khat}) in the specific case $M \simeq S$ of a 2-sphere horizon of a quiescent black hole in numerical relativity.  One striking feature of many recent numerical simulations (e.g, \cite{CLYKM:SF}) is that the horizons at late times often \textit{look} fairly regular in the fiducial spacetime coordinates used to do the evolution.  Therefore, it is natural to try a space of trial fields $\xi^a$ based simply on those coordinates.  A specific proposal follows.

Use the fiducial spacetime coordinates in which the numerical evolution occurs to induce spherical coordinates $(\theta, \phi)$ on the black-hole horizon in some more-or-less natural, but fundamentally \textit{ad hoc}, way.  Then, take the space of scalar trial potentials 
\begin{equation}\label{ThTr}
	\Theta(\theta, \phi) := \sum_{l = 1}^{l_{\mathrm{max}}}\, \sum_{m=-l}^l\, \Theta^{lm}\, \hat Y_{lm}(\theta, \phi), 
\end{equation}
where $\hat Y_{lm}(\theta, \phi)$ are the \textit{ordinary} scalar spherical harmonic functions on a \textit{round} sphere, $\Theta^{lm}$ are arbitrary constants, and $l_{\mathrm{max}}$ is a chosen cut-off.  Each of these potentials generates an area-preserving vector field via Eq.~(\ref{xipot}), and this will be our trial space 
\footnote{The space of potentials in Eq.~(\ref{ThTr}) is usually \textit{not} orthogonal, in the sense of Eq.~(\ref{xipot}), to the space of constant functions on $S$.  However, the key point is that standard properties of the ordinary spherical harmonics show that this space of trial potentials contains no actual constant functions.  This is why we have taken $l_{\mathrm{min}} = 1$ in Eq.~(\ref{ThTr}).  Therefore, Eq.~(\ref{xipot}) maps our space of trial potentials faithfully to a space of trial vector fields with the same dimension, $l_{\mathrm{max}}\, (l_{\mathrm{max}} + 2)$.}
within the full Hilbert space of Eq.~(\ref{ipdef}).  Therefore, minimize 
\begin{align}\label{Lhat}
	\ring F[\Theta] :={}& \frac{\iprod{\dual\ed\Theta}{\KL\, \dual\ed\Theta}}{\iprod{\dual\ed\Theta}{\dual\ed\Theta}} 
			= \frac{\iprod{\Theta}{\sKLs\, \Theta}}{\iprod{\Theta}{\Delta\, \Theta}} \\[1ex]\nonumber
		={}& \frac{\displaystyle \oint_S\, \bigl( 2\, g^{ac}\, g^{bd} - g^{ab}\, g^{cd} \bigr) 
				\bigl( \grad_a\, \grad_b\, \Theta \cdot \grad_c\, \grad_d\, \Theta \bigr)\, \epsilon}
			{\displaystyle -\oint_S\, \Theta \cdot \grad^a\, \grad_a\, \Theta \cdot \epsilon}
\end{align}
within the trial space of potentials given by Eq.~(\ref{ThTr}).  Generally, we should expect that the minimizing potential will generate a vector field $\xi^a_{\mathrm{trial}}$ fairly close to the minimum-eigenvalue area-preserving vector eigenfield $\ring\xi^a_{\mathrm{true}}$ of Eq.~(\ref{rKevp}).  This, in turn, should approximate Matzner's approximate Killing field from Eq.~(\ref{Kevp}).  To check the approximation, one could imagine increasing $l_{\mathrm{max}}$ until $\xi^a_{\mathrm{trial}}$ doesn't vary much with the cut-off.  Equivalently, one could use a fairly large cut-off---perhaps $l_{\mathrm{max}} = 5$ would be enough--- from the start, and check that the amplitudes $\Theta^{lm}$ are small for large $l$.  If one prefers to approximate the Cook--Whiting approximate Killing field, one need only insert a factor of the scalar curvature $R$ between the gradients in the denominator of Eq.~(\ref{Lhat}).

There is one significant issue that has not been addressed in this discussion.  Even once an approximate Killing field $\xi^a$ has been found from the eigenvalue approach, it is still determined only up to overall normalization on $S$.  For a proper rotational Killing field on a symmetric apparent horizon, the correct normalization would demand that the \textit{affine} length of each Killing orbit should be $2\pi$.  It is not immediately clear what convention might be used in the general case, without symmetry, to fix a normalization that goes over to this correct one in the limit of a symmetric manifold.  This issue will be discussed more thoroughly, in the context of practical applications, in a forthcoming paper \cite{num}.

\subsection{Acknowledgements}

The author would like to thank Ivan Booth, Manuela Campanelli, Greg Cook, Stephen Fairhurst, Greg Galloway, Carlos Lousto, Charles Torre, Bernard Whiting and Yosef Zlochower for stimulating discussions related to this question.  This work has been supported by NSF grants PHY 0400588 and PHY 0555644, and by NASA grant ATP03-0001-0027.


\begin{thebibliography}{MM}

\bibitem{K:CCL}
A.~Komar.
Covariant Conservation Laws in General Relativity.
\textit{Phys.\ Rev.}\ \textbf{113} (1959) 934-936.

\bibitem{W:GR}
R.M.~Wald.
\textit{General Relativity}.
University of Chicago Press, Chicago, 1984.

\bibitem{S:QLR}
L.B.~Szabados.
Quasi-Local Energy-Momentum and Angular Momentum in General Relativity: A Review Article.
\textit{Living Rev.\ Relativity} \textbf{7} (2004) 4.
Cited 8 July 2007.

\bibitem{ABL:IHJ}
A.~Ashtekar, C.~Beetle and J.~Lewandowski.
Mechanics of rotating isolated horizons.
\textit{Phys.\ Rev.\ D} \textbf{64} (2001) 044016.

\bibitem{AK:IDH}
A.~Ashtekar and B.~Krishnan.
Isolated and Dynamical Horizons and Their Applications.
\textit{Living Rev.\ Relativity} \textbf{7} (2004), 10.
Cited 8 July 2007.

\bibitem{BY:QLE}
J.D.~Brown and J.W.~York, Jr.
Quasilocal energy and conserved charges derived from the gravitational action.
\textit{Phys.\ Rev.\ D} \textbf{47} (1993) 1407-1419.

\bibitem{DKSS:KT}
Olaf Dreyer, B.~ Krishnan, D.~Shoemaker and E.~Schnetter.
Introduction to isolated horizons in numerical relativity.
\textit{Phys.\ Rev.\ D} \textbf{67} (2003) 024018.

\bibitem{CCGP:CKV}
M.~Caudill, G.B.~Cook, J.D.~Grigsby and H.P.~Pfeiffer.
Circular orbits and spin in black-hole initial data.
\textit{Phys.\ Rev.\ D} \textbf{74} (2006) 064011.

\bibitem{CW:S2}
G.B.~Cook and B.F.~Whiting.
Approximate Killing Vectors on $S^2$.
E-Print \texttt{arXiv:~0706.0199v1~[gr-qc]}.  2007.

\bibitem{M:ASS}
R.A.~Matzner.
Almost Symmetric Spaces and Gravitational Radiation.
\textit{J.\ Math.\ Phys.}\ \textbf{9} (1968) 1657-1668.

\bibitem{num}
C.~Beetle, M.~Campanelli, C.O.~Lousto and Y.~Zlochower.
In preparation.

\bibitem{Y:DST}
J.W.~York, Jr.
Covariant decompositions of symmetric tensors in the theory of gravitation.
\textit{Ann.\ Inst.\ Henri Poincar\'e} \textbf{21} (1974) 319-332.

\bibitem{YO:IVP1}
N.~\'O~Murchadha and J.W.~York, Jr.
Initial-value problem of general relativity.  I.  General forumlation and physical interpretation.
\textit{Phys.\ Rev.\ D} \textbf{10} (1974) 428-436.

\bibitem{CLYKM:SF}
M.~Campanelli, C.O.~Lousto, Y.~Zlochower, B.~Krishnan and D.~Merritt.
Spin flips and precession in black-hole-binary mergers.
\textit{Phys.\ Rev.\ D} \textbf{75} (2007) 064030.

\bibitem{MW:MMP}
J.~Mathews and R.L.~Walker.
\textit{Mathematical Methods of Physics}.
Addison-Wesley, Redwood City, California, 1970.

\end{thebibliography}
\end{document}